\documentclass[structabstract]{aa}
\usepackage{txfonts}
\usepackage{graphicx}
\usepackage{bm}
\usepackage{natbib}
\usepackage{color}

\newcommand{\rlc}{r_{\rm LC}}
\newcommand{\blc}{B_{\rm LC}}
\newcommand{\ghat}{\hat{\gamma}}
\newcommand{\alc}{a_{\rm LC}}

\newcommand{\Gwind}{\Gamma}
\newcommand{\Edot}{\dot{E}_{33} }
\newcommand{\Rmin}{R_{\rm min}}
\begin{document}

	\title{Pulsed high energy $\gamma$-rays from thermal populations in the current sheets of pulsar winds}

	\author{I. Arka 
	\and 
	G. Dubus}

	\institute{Institut de Plan\'etologie et d' Astrophysique de Grenoble (IPAG), UMR 5274, 38041 Grenoble, France
	\email{arkai@obs.ujf-grenoble.fr}}

    \date{Received ; Accepted }

\abstract{ More than one hundred GeV pulsars have been detected up to now by the LAT telescope on the {\em Fermi} gamma-ray observatory, showing peak energies around a few 
GeV. Current modelling proposes that the high energy emission comes from outer magnetospheric gaps, however radiation from the equatorial current sheet which separates the two magnetic 
hemispheres outside the light cylinder has also been investigated.} 
{In this paper we discuss the region right outside the light cylinder, or "near wind" zone. We investigate the possibility that synchrotron radiation emitted by thermal 
populations in the equatorial current sheet of the pulsar wind in this region can explain the lightcurves and spectra observed by {\em Fermi}/LAT.}
{ We use analytical estimates as well as detailed numerical computation to calculate the $\gamma$-ray luminosities, lightcurves and spectra of $\gamma$-ray pulsars. }
{ Many of the characteristics of the $\gamma$-ray pulsars observed by {\em Fermi}/LAT can be reproduced by our model, most notably the position of these objects in the
$P-\dot{P}$ diagram, and the range of $\gamma$-ray luminosities. A testable result is a sub-exponential cutoff with an index $b=0.35$. We also predict the existence of a population of pulsars 
with cutoff energies in the MeV range. These have systematically lower spindown luminosities than the {\em Fermi}/LAT detected pulsars. }
{ It is possible for relativistic populations of electrons and positrons in the current sheet of a pulsar's wind right outside the light cylinder to emit synchrotron radiation that
peaks in the sub-GeV to GeV regime, with $\gamma$-ray efficiencies similar to those observed for the {\em Fermi}/LAT pulsars.}


\titlerunning{Pulsed $\gamma$-rays from thermal particles in pulsar wind current sheets}
\authorrunning{I. Arka \& G. Dubus}
   \maketitle

\section{Introduction}

Since the launch of the {\em Fermi} space telescope, the sample of gamma ray pulsars has grown to include more than one hundred objects. All these pulsars, 
whether young or millisecond, single or in binaries, show spectra consistent with power laws with exponential cutoffs, where the cutoff 
energy lies in the range 1-10 GeV \citep{latfirstyear}. A prominent exception is the Crab pulsar, which has been detected by ground-based 
Cerenkov arrays in the TeV regime, and the spectrum of which can be well fitted by a broken power law \citep{crabmagic,mccannetal11}. These observations point 
to the outer gap/slot gap models as the most probable explanation for the pulsed gamma-ray emission, based on both population prediction statistics and 
lightcurve modelling \citep{gonthieretal10,decesaretal11,pierbattistaetal11,venteretal11,watters+romani11,venteretal12}, although for millisecond pulsars 
one has to evoke non-dipolar field geometries or displaced polar caps, in order to reach the required energies \citep{harding+muslimov11}. In those models, the 
emission originates within the light cylinder, defined by the cylindrical radius $\rlc = c/\omega$ from the pulsar's rotational axis, where a corotating particle would 
reach the speed of light. However it was pointed out by \cite{bai+spitkovsky10} that the emission region must be extended slightly outside the light cylinder 
in order to reproduce the doubly-peaked light curves using the magnetic field configuration from self-consistent, force-free simulations of pulsar magnetospheres.

The idea that high energy pulsations might come from current sheets near or outside the light cylinder is not new. \cite{lyubarsky96} predicted that particles 
accelerated though reconnection close to the light cylinder might emit gamma-rays through the synchrotron process. Lyubarsky's emission site is the 
point where the warped equatorial current sheet which separates regions of opposite magnetic field polarity in the pulsar's wind meets the current flowing 
to/from the pulsar's polar caps, the so-called Y-point \citep{spitkovsky06}. Later it was pointed out that pulsed emission naturally arises from the periodicity of the 
pulsar wind, which is modulated by the star's rotation, in combination with the large bulk Lorentz factors of the outflow \citep{kirketal02}. This model has been 
successfully applied to calculate the polarization of optical emission from the Crab pulsar \citep{petri+kirk05} and it has also shown promise in explaining the 
gamma-ray light curves observed by {\em Fermi}/LAT \citep{petri11}. In these models the emission is 
attributed to the inverse Compton process and starts in the wind region far from the light cylinder, $r \gg \rlc$. In the same context it was proposed that 
the gamma-ray radiation of the {\em Fermi}/LAT band could be synchrotron radiation from power-law electrons in the pulsar wind's equatorial sheet far from the light cylinder 
\citep{kirketal02,petri12}.

There is, however, no obvious reason why emission should be truncated outside the light cylinder (for magnetospheric gap models) or start at a minimum radius 
in the far wind zone (for the striped wind model). The emission region should evolve continuously from the outer magnetospheric gaps through the light cylinder and 
into the  equatorial current sheet, possibly affected by the reconnection process that will inevitably occur in the sheet \citep{bai+spitkovsky10}. The region beyond 
the light cylinder but still in the "near wind" zone is the area that we are trying to explore. The purpose of this paper is to demonstrate how emission in the 
100~MeV-100~GeV range explored by {\em Fermi}/LAT can naturally arise from thermal populations of particles in the striped wind, without 
going into the details of reconnection physics, but just using a few simple assumptions about the local description of the current sheet. In section 2 we describe 
the model we use, including our assumptions, and give some analytical estimates for the emitted radiation. In section 3 we present examples of 
emission maps and phase-averaged spectra which arise from our model and explore the parameter space in order to come to more general conclusions about the
pulsar population observed by {\em Fermi}/LAT. Finally we discuss our 
results and the potential of a more detailed description of the current sheet outside the light cylinder as a source of pulsed energetic radiation.

\section{Equatorial current sheet}

\subsection{Description of the particle population}

In order to describe the wind outside the light cylinder, we will use the "slow-rotator" solution to the magnetohydrodynamics equations governing the physics of the 
pulsar wind, which was found by \cite{bogovalov99}. This solution refers to an electron-positron wind launched by a rotating neutron star, the magnetic axis of 
which is at an angle $\chi$ to its rotational axis. In this description the wind is purely radial and super-fast magnetosonic at launch, and the field can be 
described by a radial and an azimuthal component of respective magnitude:
\begin{eqnarray}
B_r  &=& \frac{\blc}{R^2} \\
B_{\varphi} &=&  \frac{\blc}{\beta R} \sin\vartheta
\end{eqnarray}where $R=r/\rlc$ is the spherical radius normalized to the light cylinder, $\vartheta$ is the polar angle, $\beta = (1-1/\Gamma^2)^{1/2}$ is the bulk speed
of the wind, normalized to the speed of light and $\blc$ is a fiducial magnetic field magnitude at the light 
cylinder. The adequacy of this solution in describing approximately the structure of the wind for radii as small as the light cylinder has been confirmed by force-free simulations of pulsar 
magnetospheres \citep{spitkovsky06,bai+spitkovsky10}.

A prominent feature of Bogovalov's solution is the warped equatorial current sheet, separating the two magnetic hemispheres, across which the field changes
sign. In the mathematical solution the sheet is just a discontinuity, however in reality it should have a finite thickness. Such a current sheet is populated by hot particles, 
the pressure of which balances the magnetic pressure of the cold, strongly magnetized plasma outside the sheet. The sheet oscillates in space and time with a 
wavelength of $\lambda = 2 \pi \beta/\omega$, and with the pulsar's frequency 
$\omega = 2\pi/P$, where $P$ is the pulsar's period measured in seconds.

We will assume that, in the wind frame (the frame that propagates radially outwards with a Lorentz factor equal to the bulk Lorentz factor of the wind, 
$\Gamma$), the sheet can be locally described by the relativistic Harris equilibrium \citep{hoh66}. For this description to be valid, the segment of the sheet under 
consideration should be approximately flat. For a relativistic outflow, the hydrodynamically causally connected region of the flow has an opening ange of roughly $\sim 
1/\Gamma$, centered on the direction of motion of the outflow. Since for a pulsar wind $\Gamma \gg 1$, the hydrodynamically connected sheet segment can be considered
flat and the local Harris sheet description should be a good approximation.

We will denote quantities measured in the wind frame with a prime. If we denote the sheet normal direction in the wind frame with a capital $X'$, and the 
sheet midplane is at $X'=0$, then locally the field inside and outside the segment can be described by a tangent hyperbolic profile: 
\begin{equation}
B'(X') = \pm B'_0 \tanh\left( \frac{X'}{\delta'} \right) 
\end{equation}where $B'_0$ is the magnitude of the field outside the sheet, $\delta'$ is a measure of the sheet thickness and the field is in the direction parallel to the 
sheet and perpendicular to the direction of current flow, which is locally considered to be $Z'$. The particle population in the sheet consists of two  
counter-drifting relativistic Maxwellian distributions the density of which falls with $X'$, and whose drift provides the net current. From the pressure balance across 
the sheet follows the dependence of the density of each distribution on $X'$ \citep{kirk+skjaeraasen03}:
\begin{eqnarray}
N'_{\pm} & = & N'_{\pm 0} \cosh^{-2} \left( \frac{X'}{\delta'}\right) \\
N'_{\pm 0} & = & \frac{B'^2_0}{16 \pi m c^2 \Theta}
\label{density}
\end{eqnarray}where $\Theta = k_B T'/(mc^2)$ is a dimensionless temperature associated with the particle distribution. The condition connecting the particle density, 
temperature and current sheet thickness is \citep{kirk+skjaeraasen03,lyubarsky+kirk01}:
\begin{equation}
\frac{\Theta m c^2}{4 \pi N'_{\pm 0} e^2 \gamma_{\pm}} = \delta'^2 \beta^2_{\pm}
\label{sheetthickness}
\end{equation}where $\gamma_{\pm}$ and $\beta_{\pm}$ are the Lorentz factor and the corresponding speed, normalized to the speed of light, of the drift of the 
distributions, in the wind frame. Finally, by assuming that the ideal gas law holds for the distribution we get:
\begin{eqnarray}
p' = 2 N'_{\pm} T' = (\ghat -1)(e'-2N'_{\pm}mc^2)
\end{eqnarray}where $e'$ is the energy density associated with the two counter-streaming distributions. For the relativistic Harris sheet solution, this equation 
becomes:
\begin{equation}
\ghat = 1 +\frac{\Theta}{\gamma^2_{\pm} \left(1+\beta^2_{\pm} \right) \Theta + \gamma_{\pm} \left( K_1(1/\Theta)/K_2(1/\Theta) -1\right)}
\label{ghat}
\end{equation}where $K_1$ and $K_2$ are modified Bessel functions of the second kind. In the inner part of the pulsar wind, which is of interest to us, the 
thermal particles in the distribution are predicted to be highly relativistic, something which allows us to set $\ghat =4/3$ in the above equation and also
has as a result that the thermal energy of the particles in the distributions will greatly exceed their rest-mass energy: $\Theta \gg 1$. Using the 
approximations
\begin{eqnarray}
K_1(x \ll 1) &\simeq & x^{-1} \\
K_2(x\ll 1) &\simeq & 2 x^{-2}
\label{k2approx}
\end{eqnarray}Eq.~\ref{ghat} can be simplified to:
\begin{equation}
\beta_{\pm} \simeq \frac{1}{\sqrt{\Theta}} \enspace .
\end{equation}
This means that the drift of the distributions in the current sheet is not relativistic, and beaming effects caused by it can be ignored when calculating the emitted radiation.
Therefore the distributions can be for this purpose considered to be isotropic in the frame in which the sheet is at rest. Using this approximation we can solve 
Eqs.~\ref{density} and \ref{sheetthickness} for $\Theta$ and $N'_{\pm 0}$. In doing this, we have to take into account the fact that the angle $\arccos( \hat{n}\cdot\hat{r})$ 
between the sheet normal and the radius unit vector changes when moving from the lab to the wind frame.

The magnetic field from the solution of \cite{bogovalov99} is always parallel to the sheet, if one ignores reconnection effects. This means that close to the light 
cylinder and for a radial outflow one cannot ignore the radial component of the field. The full field {\em in the wind frame} will be:
\begin{equation}
B'_0 = \frac{\blc}{R^2} \sqrt{1+\left(\frac{R\sin\vartheta}{\beta\Gamma} \right)^2}
\end{equation}and the toroidal component will prevail in this frame only for $R > \Gamma$ (we will call this region the "far wind" region). Conversely, in the region 
$R < \Gamma$ (which we will call the "near wind" region), it is
a good approximation to ignore the second term under the square root and approximate $B'_0 \simeq \blc/R^2$.

If one considers a perpendicular rotator (a pulsar with $\chi=\pi/2$), the temperature and density in a current sheet segment are given by the simple expressions:
\begin{eqnarray}
\Theta  &=& \left( \frac{\alc \Delta \sin\vartheta}{2R} \right)^{2/3} \label{theta}\\
N'_{\pm 0} &=& \frac{B'^2_0}{16\pi mc^2\Theta}
\end{eqnarray}
The dimensionless parameter $\alc$ appearing above is called the strength parameter, and is defined at the pulsar's light cylinder as:
\begin{equation}
\alc = \frac{e \blc P}{2 \pi m c} \enspace .
\end{equation}Taking a pulsar's moment of inertia to be $I=10^{45} \, \rm g cm^2$, the spindown luminosity is calculated by the period and the 
period derivative as \citep{latfirstyear}:
\begin{equation}
\Edot = 4\pi^2 10^{12} \dot{P} P^{-3}
\end{equation}where we have normalized the luminosity to the value $10^{33}$ erg/s. The magnetic field (in Gauss units) can then be expressed as: 
\begin{equation} 
\blc = 46.83\Edot^{1/2}P^{-1}
\label{fieldlc}
\end{equation}
The strength parameter then can be expressed as a function of the spindown luminosity only:
\begin{equation}
\alc = 1.3\times 10^8\Edot^{1/2}
\end{equation}
All GeV pulsars detected so far have strength parameters in the range $10^8-10^{11}$.

Finally, $\Delta <1$ is the fraction of a half wavelength that a sheet occupies, measured in the {\em radial} direction (i.e. not perpendicularly to the local sheet plane), 
in the lab frame. We will assume that $\Delta$ is constant 
with radius for the sake of simplicity, however it is far from clear how this parameter evolves with radius and obliquity 
$\chi$ close to the light cylinder. We will also assume that 
$\Gamma$ is constant, although generally reconnection in the current sheet has been shown to increase $\Gamma$ with radius in the far wind zone
\citep{lyubarsky+kirk01,kirk+skjaeraasen03}, a result that might also apply to the near wind zone. These assumptions are based on the fact that only a very 
limited radius interval contributes to the gamma-ray radiation, as we will see below, therefore if the change in $\Delta$ and $\Gamma$ happens with a scale larger 
than $r_{\rm LC}$, it is not relevant to the present estimations.

Substituting for $\alc$ in Eq.~\ref{theta} we get:
\begin{equation}
\Theta = 1.6\times 10^5 \Edot^{1/3} \left( \frac{\Delta\sin\vartheta}{R}\right)^{2/3} 
\label{temperature}
\end{equation}The requirement that $\Theta \gg 1$ in the near wind region $R<\Gamma$ translates to:
\begin{equation}
\Delta \gg 10^{-8}\frac{\Gwind}{\Edot^{1/2}\sin\vartheta}
\label{thetalarge}
\end{equation}which, as we will see in the following, is easily satisfied for all $\gamma$-ray pulsars, provided the emitting region is not very close to the polar axis. 
In any case, the validity of 
the $\Theta \gg 1$ assumption has to be checked a posteriori for all cases of studied pulsars.

In the following, all results will refer to the near wind region, $R<\Gamma$, but still outside the pulsar's light cylinder, close to which we assume that the current sheet is formed. Therefore we will 
consider a minimum value for the radius of the radiating sheet of $\Rmin =1$. Our model is valid only for relativistic outflows, so we will also set a lower limit on $\Gamma$ of $\Gamma_{\rm min} 
=10$.

In the next paragraphs, we assume a perpendicular rotator ($\chi = \pi/2$) in order to give some analytical estimates concerning the emitted radiation and the characteristics of the
current sheet and its populations. These estimates can be generalized to the oblique rotator (see Appendix~\ref{apppend2}), however the conclusions remain essentially the same as long as 
the line of sight is not close to the edge of the current sheet, where $\zeta = \pi/2- \chi$.
 
\subsection{Larmor radius of hot particles}

An important condition for the consistency of our model comes from the requirement that the hot particles' Larmor radius in the full field between the current 
sheets should be smaller than the sheet width. In the wind frame this requirement translates to:
\begin{equation}
\frac{\langle \gamma \rangle m c^2}{e B'_0} \gg \Delta \rlc R \left( 1+ \frac{R^2}{\Gamma^2}\right)^{-1/2} \enspace .
\end{equation}
This inequality is equivalent to $\Theta^{1/2} \gg 1$ which should hold close to the light cylinder, by the assumptions of the model.
 
\subsection{Energy of emitted photons}

The mean Lorentz factor of the electron/positron distribution in the current sheet is $\langle \gamma \rangle \sim 3\Theta$. Electrons of this $\langle \gamma 
\rangle$ gyrating in the field $B'_0$, radiate photons of energy (in the observer's frame):
\begin{equation}
\mathcal{E} \approx  \frac{3}{2} \Gamma (3\Theta)^2  \hbar \frac{eB'_0}{mc}
\end{equation}
After some manipulation, one finds the highest observable photon energy:
\begin{equation}
\mathcal{E}_{\rm GeV} \approx 2 \times 10^{-4} \frac{\Edot^{7/6}}{P} \frac{\Gamma (\Delta\sin\zeta)^{4/3}}{R_{\rm min}^{10/3}}
\label{cutoff}
\end{equation}where the photon energy is measured in GeV. In Eq.~\ref{cutoff} we have kept only the poloidal component of the magnetic field, since it dominates close to the light
cylinder, at $R < \Gamma$. All pulsars in the first {\em Fermi}/LAT catalogue have $\Edot^{7/6}/P\gg 1$. This is why it is possible for
most objects in the catalogue to find an appropriate combination of $\Gamma$, $\Delta$ and $R_{\rm min} \geq 1$ which will bring $
\mathcal{E}_{\rm GeV}$ to the regime observed by 
LAT, while at the same time satisfying all the restrictions on $\Delta$, $\Gamma$ and $R$. This is especially easy for the highest luminosity pulsars
($\Edot \gg 1$), or 
for the very low period ones ($P\ll 1$ s, millisecond pulsars), as one can deduce by inspecting Eq.~\ref{cutoff}. In other words, for pulsars of the same
spindown luminosity, millisecond pulsars are more likely to reach GeV energies. Alternatively, if $\Gamma$ and
$\Delta$ are the same, millisecond pulsars with lower spindown luminosities than young pulsars can emit in the {\em Fermi}/LAT band, a trend that might be observable as our statistics
on gamma-ray pulsars increase.

From the constraints on $\Delta$ and $R$, we can calculate an absolute maximum on the peak energy of the emitted spectrum, for $\zeta = \pi/2$:
\begin{equation}
 \mathcal{E}_{\rm GeV,max}\simeq 2.3 \times 10^{-5} \Gamma \frac{\Edot^{7/6}}{P}
\label{max2}
\end{equation}
As we will see below, $\Delta$ is often more severely constrained by the radiation reaction limit, so the energy $\mathcal{E}_{\rm GeV,max}$ will not necessarily be 
reached for all objects.

The rapid fall of the peak photon energy with radius means that in this simple model the most energetic photons come from close to the 
light cylinder. In a realistic situation it is to be expected that either $\Gamma$ or $\Delta$ or both will rise with radius, making the fall of $\mathcal{E}_{\rm GeV}$ 
with radius less steep, however even if $\Delta$ and $\Gamma$ proved to rise linearly with $R$, the emitted photon energy would still fall as $\propto R^{-1}$.

\subsection{Energy losses}

For the relativistic Harris equilibrium to hold in a quasi-steady state, equilibrium has to be established in the wind frame within a 
timescale of the order of magnitude $t'_{\rm R} \sim R(\Gamma \omega)^{-1}$, which is the timescale on which the magnetic field change is comparable to its 
magnitude $\delta B' \sim B'$. Equilibrium is established in the current sheet within a timescale $$t'_{\rm eq} \lesssim \frac{\Delta \rlc}{c}$$Therefore the requirement 
$t'_{\rm eq} < t'_{\rm R}$ holds close to the light cylinder as long as $\Delta \lesssim\Gamma^{-1}$, while this condition is relaxed with radius to $\Delta \lesssim R/\Gamma$.

For the particles not to suffer catastrophic energy losses, their synchrotron cooling timescale $t'_{\rm s}$ should be larger than 
$t'_{\rm R}$. Comparing the two one finds:
\begin{eqnarray}
\frac{t'_{\rm s}}{t'_{\rm R}} \simeq 4.57 \frac{P}{\Edot^{4/3}} \frac{\Gamma R^{11/3}}{(\Delta\sin\zeta)^{2/3}} \enspace .
\label{tratio}
\end{eqnarray}For the lower $\Edot$ pulsars the inequality $t'_{\rm s}\geqslant t'_{\rm R}$ generally holds for $R > \Rmin$, but this may not be the case for some millisecond
pulsars, because of the linear dependency of the ratio on pulsar period $P$. The steep radius dependence, however, relaxes this condition rapidly, so that when it breaks 
down, it does so only for a very limited $R$-range close to the light cylinder.

If $t'_{\rm s}/t'_{\rm R} \leqslant 1$, as is often the case for high spindown pulsars and millisecond pulsars, particles will lose energy rapidly and cool, disturbing the 
pressure equilibrium between the populations in the current sheet and the magnetic field outside it. The pressure in the sheet will fall, thus causing its compression by the 
external magnetic field. This can initiate compression driven magnetic reconnection, a phenomenon previously studied in the context of the interaction of current sheets in a pulsar wind 
with the wind's termination shock \citep{lyubarsky03,petri+lyubarsky07,lyubarsky+liverts08,sironi+spitkovsky11}. Reconnection has been shown to be able to accelerate particles to non-
thermal distributions above 
the thermal peak of the particle spectrum (and also, as already mentioned, to accelerate the bulk flow and cause the sheet width to rise). A detailed study of this radiation loss induced 
process is beyond the scope of this work. It should be stressed, however, that if $t'_{\rm s}/t'_{\rm R} <1$, some 
acceleration processes ought to be at work in order to supply the sheet with energetic particles. This will in all likelihood result in power-law distributions in 
the current sheet, a signature that might be observable in the spectrum above the MeV-GeV peak for the highest luminosity pulsars and for those millisecond pulsars with
$t'_{\rm s}/t'_{\rm R} \leqslant 1$ (something that has not been modelled in the present article).

\begin{figure*}\begin{center}\begin{minipage}{0.5\textwidth}
\includegraphics[scale=1]{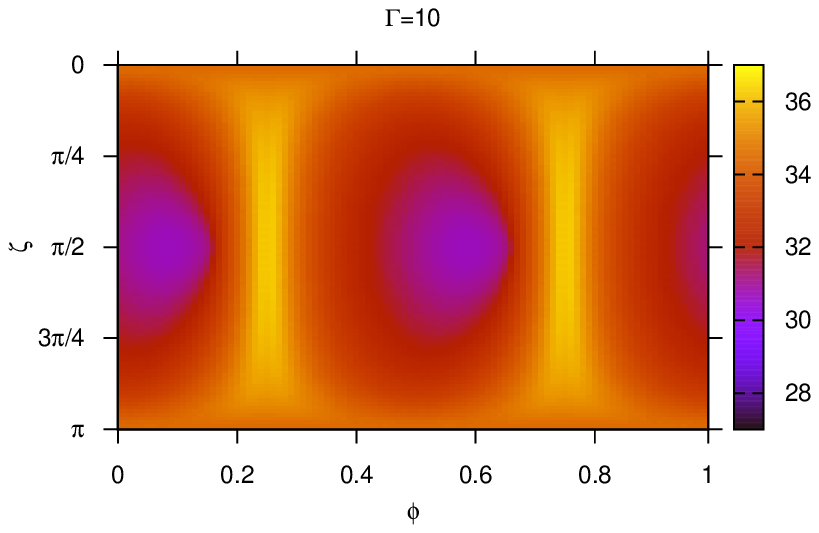}
\includegraphics[scale=1]{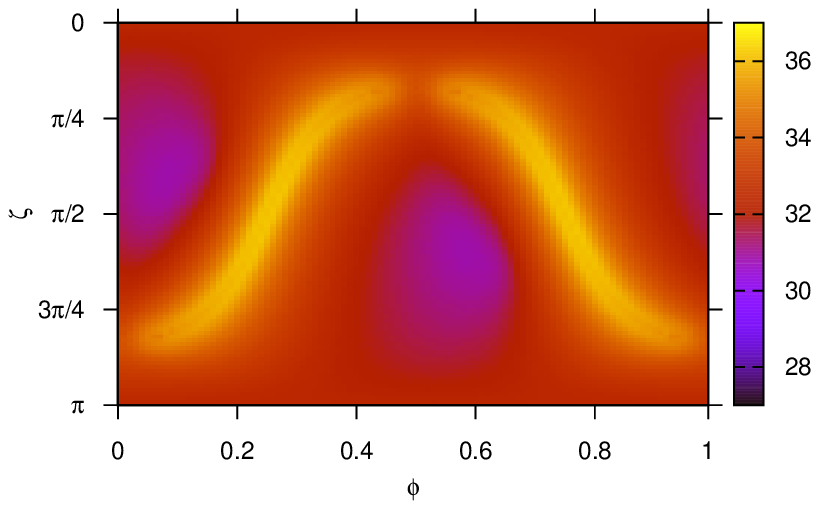}
\includegraphics[scale=1]{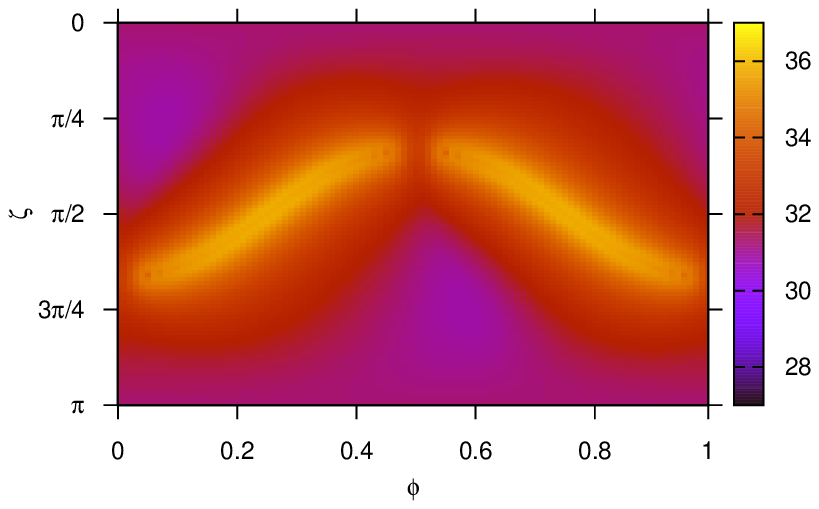}
\end{minipage}\begin{minipage}{0.5\textwidth}
\includegraphics[scale=1]{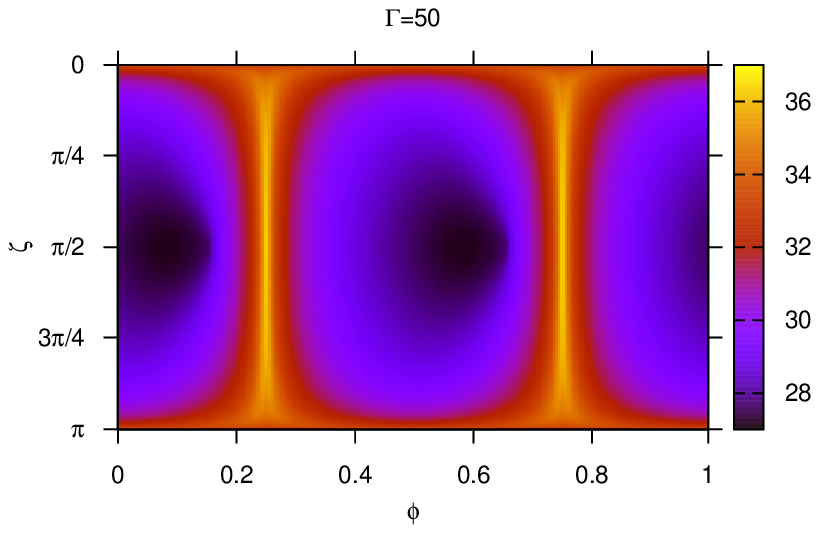}
\includegraphics[scale=1]{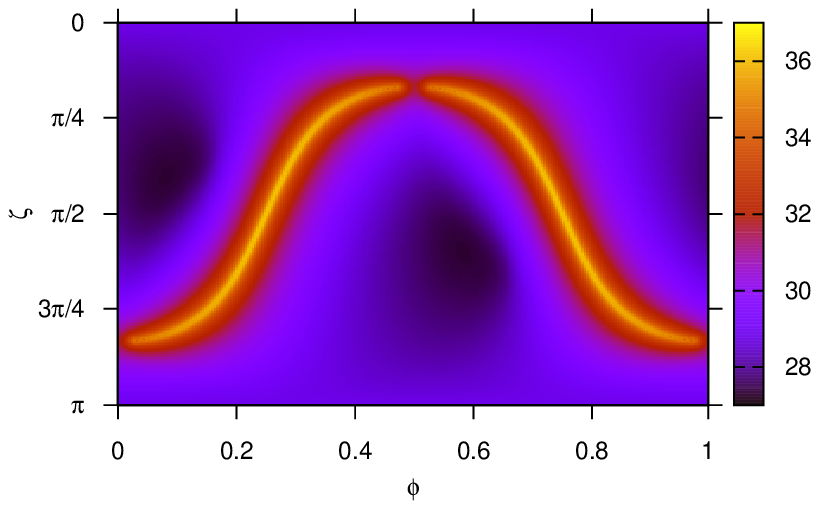}
\includegraphics[scale=1]{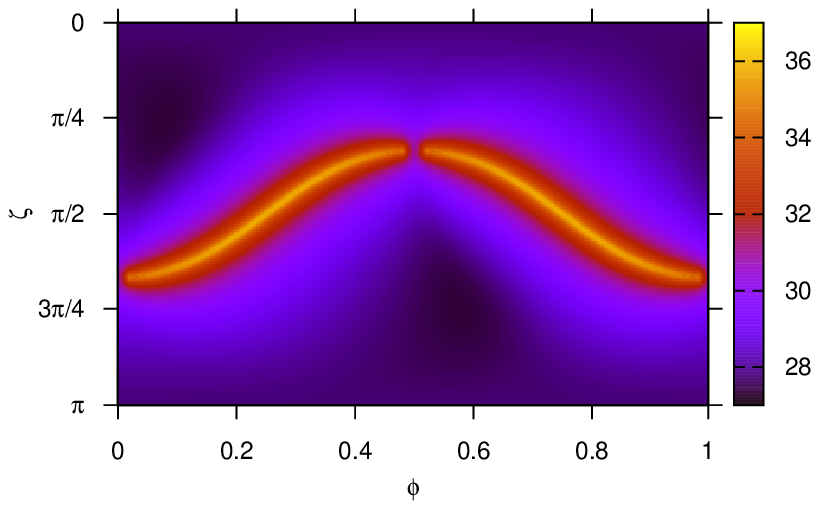}\end{minipage}
\end{center}
\caption{Sky maps of the pulsar's total luminosity shown for two different bulk Lorentz factors $\Gamma=10$ and $\Gamma = 50$ and for different $\chi$. In the upper panels $\chi = 
\pi/2$, in the middle ones $\chi = \pi/3$ and in the lower ones $\chi= \pi/6$. All the plots have been made using the parameters $\Delta = 0.01$ and $R_{\rm min} =1$. The luminosity
is given by integrating $\mathcal{L}_{\nu}$ (Eq.~\ref{luminosity}) over all frequencies and multiplying with $d^2$, where $d$ is the distance of the pulsar from the observer. A beaming factor 
of $\mathcal{F}_{\Omega}= (4\pi)^{-1}$ has been used.}
\label{sample}
\end{figure*}

\subsection{Radiation reaction limit and maximum emitted energy}

If the particles of mean energy $\sim 3\Theta mc^2$ in the current are gaining energy through acceleration in an electric field $E\sim \xi B$ with $\xi<1$, then their 
acceleration is limited by radiation losses. For radiation-reaction limited synchrotron emission (in the full field amplitude $B'_0$), energy losses compensate
possible energy gain:
\begin{equation}
\left[\frac{dW}{dt}\right]_{\rm syn} = e\xi B'_0 c \leqslant e B'_0 c
\end{equation}From this expression one can calculate an upper limit for $\Delta$:
\begin{equation}
\Delta \leqslant \Delta_{\rm lim} = 256 \frac{R^{5/2} P^{3/4}}{\Edot^{7/8}\sin\zeta}
\end{equation}
The $\Edot$ dependence in the denominator implies that for the most powerful pulsars, the sheet is thinner close to the light cylinder. For the Crab we get 
$\Delta_{\rm lim}=3\times 10^{-4} \sin^{-1}\zeta$ at $R_{\rm min}$, whereas for most of the weaker pulsars with $\Edot \leq 10^2$, $\Delta_{\rm lim}$ is greater than 
unity, therefore not presenting a constraint. All millisecond pulsars detected by {\em Fermi}/LAT fall into this category.

Inserting $\Delta_{\rm lim}$ in Eq.~\ref{cutoff} one recovers the classical synchrotron limit:
\begin{equation}
\mathcal{E}_{\rm GeV,max} < 0.36 \Gamma
\label{max1}
\end{equation}where the factor $\Gamma$ comes from the boosting of the photon's energy to the lab frame.
The minimum of Eqs.~\ref{max2},\ref{max1}, then, can give an estimate of the maximum value of the GeV cutoff for any individual object.

Inserting $\Delta_{\rm lim}$ into Eq.~\ref{tratio}, one gets a lower limit for the ratio $t'_{\rm s}/t'_{\rm R}$ (applicable mainly to higher-spindown 
pulsars, for which $\Delta_{\rm lim}<1$):
\begin{equation}
\frac{t'_{\rm s}}{t'_{\rm R}} > \left[ \frac{t'_{\rm s}}{t'_{\rm R}}\right]_{\rm min} = 0.23 \Gamma  R^2 P^{1/2} \Edot^{-3/4}
\end{equation}
We see that, mainly for young pulsars of large spindown luminosity, the right hand side can be less than unity. These pulsars either have smaller $\Delta$ in order to 
keep $t'_{\rm s}/t'_{\rm R}>1$, or rapid reacceleration mechanisms in their current sheets as argued above (or both). Therefore there should be a  trend in the $P-\dot{P}$
diagram to observe more objects with acceleration signatures (i.e. power-law tails) in their spectrum as one moves to higher spindown luminosities, or equivalently as one moves to higher 
$\dot{P}$ and lower $P$ (upward left region in a $P-\dot{P}$ diagram).

\section{Lightcurves, spectra and $\gamma$-ray luminosity}

\subsection{Lightcurves}

The expected lightcurves and spectra produced in the current sheets of a pulsar can be computed numerically by integrating the emission coefficient of the particles in the magnetic field of the 
current sheet along the line of sight to the observer. The procedure is explained in the Appendix.

In the wind model of the pulsar radiation two pulses per period are expected, the separation and the width of which varies with the obliquity $\chi$ and the line of sight
angle $\zeta$. The width of the pulses depends on the bulk speed of the outflow, with wider pulses for lower $\Gamma$ \citep{kirketal02,petri11}. Examples of the variations of 
lightcurves with $\chi$, $\zeta$ and $\Gamma$ are shown in Fig.~\ref{sample}, where we have used a model pulsar of surface magnetic field equal to $B= 10^{10}$G and period $P=0.01$ 
s in order to plot the sky maps of the pulsed total luminosity.

In the left column of Fig.~\ref{sample} sky maps are shown for $\Gamma = 10$, while in the right column the value of the bulk Lorentz factor is $\Gamma=50$. The 
width of the pulse depends strongly on $\Gamma$, as expected. 

The upper maps in both columns represent the case of the perpendicular rotator, where $\chi = \pi/2$, the middle ones are for $\chi = \pi/3$ and the lower ones for $\chi=\pi/6$.
As the obliquity falls, the width of the pulse increases while the peak luminosity falls, resulting in the overall phase-averaged luminosity remaining essentially constant for different obliquities. 
The widest pulses are observed for the smallest inclination angles. However, for very small $\chi$ the line of sight to the observer has to lie very close to the equatorial plane of the pulsar for 
the pulse to be observed, since the luminosity is significant only for $\zeta \leq \pi/2 - \chi$. This is a likely geometry for the objects which are observed to have two wide pulses with a phase 
separation of $\delta \phi \sim 0.5$. 

From the sky maps in the case of $\Gamma=10$ it can be discerned that there is a slight substructure in the pulses. There is a slight dip in luminosity at the peak of each pulse, resulting in 
two sub-peaks appearing for all $\zeta$. This is caused by the structure of the current sheet. In the middle of the sheet the magnetic field is exactly zero, rising towards the sides, while the density 
of the hot particles is highest and falls towards the sides. Therefore the bulk of the radiation of each 
current sheet segment comes from two regions away from the sheet midplane, where the product of particle density and magnetic field is largest. The double-peaked shape of the pulse
reflects the two luminosity peaks within the sheet. Because of the shrinking
pulse thickness with rising $\Gamma$ this phenomenon should be observable only for the lowest bulk Lorentz factors.

\begin{figure}
\includegraphics[scale=1]{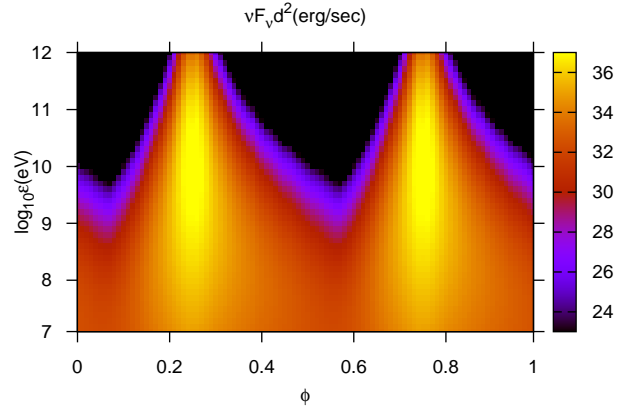}
\caption{The peak flux $\nu F_{\nu}$(multiplied by the square of the distance to give a luminosity estimate) as a function of phase $\phi$ and peak energy $\mathcal{E}$. $\mathcal{E}$ ranges from 
10 MeV to 1 TeV. The same pulsar parameters were used as in Fig.~\ref{sample}, with $\chi = \pi/3$ and $\zeta = \pi/2$.}
\label{Emaxmap}
\end{figure}

It is useful to note here that the highest luminosities in the sky maps are correlated with the highest peak energies $\mathcal{E}_{\rm GeV}$, resulting in pulses that get sharper for higher 
energy photons. This is shown in Fig.~\ref{Emaxmap}, where the flux $\nu F_{\nu}$ has been plotted in a color map as a function of emitted energy $\mathcal{E}$ (in eV) and phase $\phi$. A 
horizontal slice of the map shows the lightcurve at a specific energy while a vertical slice shows the spectrum at a given phase. The pulses become narrower with increasing energy, as
can be seen from the wider flux variation in a horizontal slice as one moves higher in the map. Also, the peak energy varies by many orders of magnitude within one phase, and is lowest 
between pulses, something that can be seen when comparing spectra at different phases.

\begin{figure}
\includegraphics[scale=1]{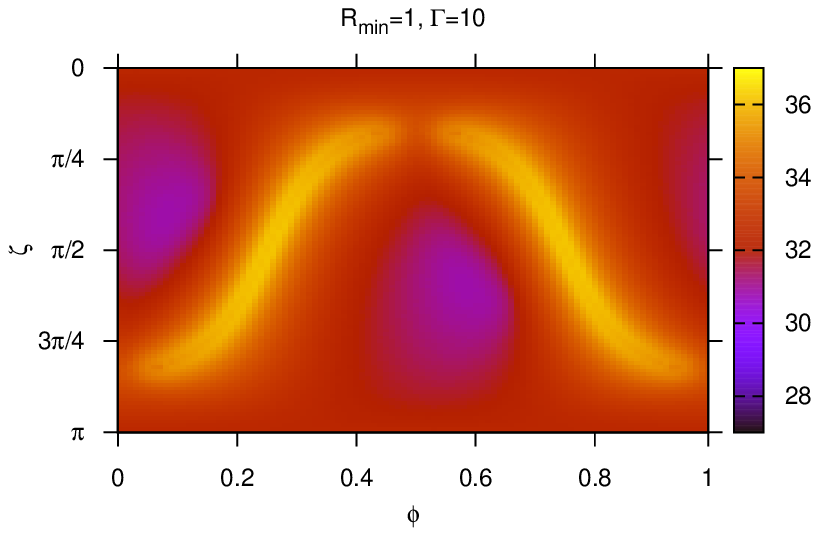}
\includegraphics[scale=1]{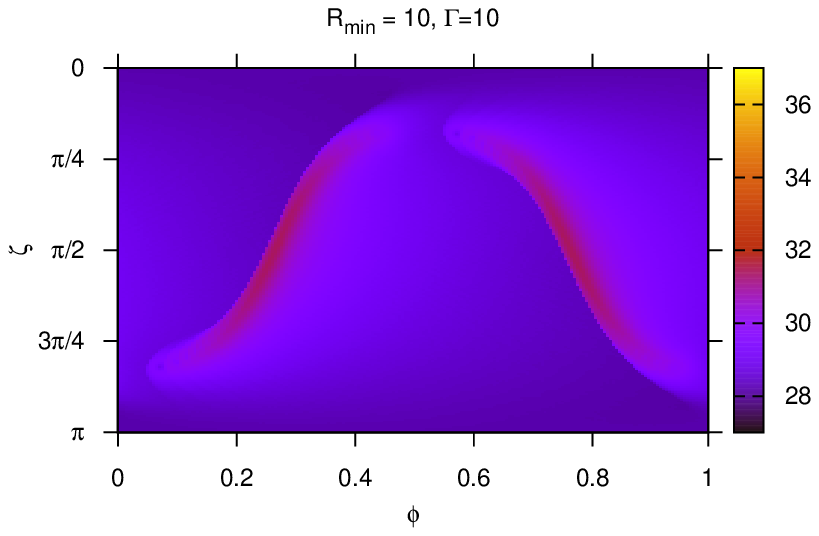}
\includegraphics[scale=1]{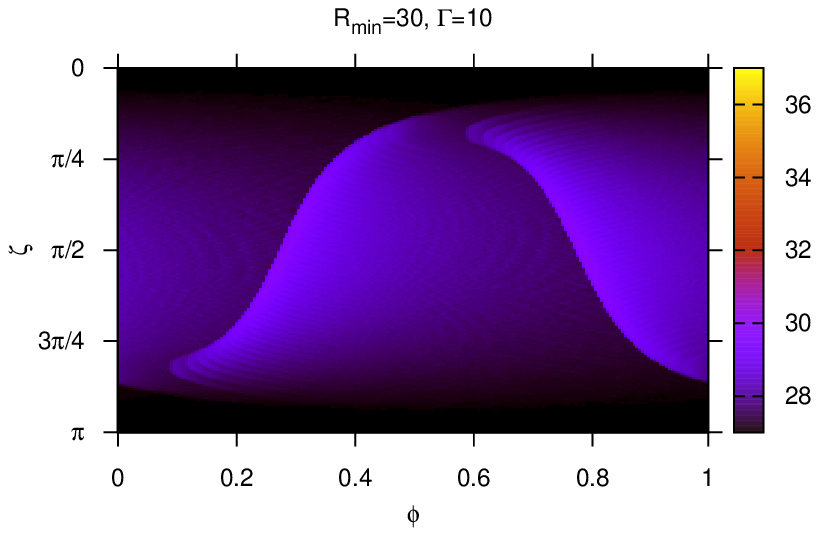}
\caption{Variation of the pulsar's lightcurves for different $R_{\rm min}$. The bulk Lorentz factor in these examples is $\Gamma=10$ and
the obliquity is $\chi = \pi/3$. The rest of the parameters are the same as in Fig.~\ref{sample}.}
\label{Rmin}
\end{figure}

In Fig.~\ref{Rmin} we show how the lightcurves change when moving the minimum radius $R_{\rm min}$ from the light cylinder to a distance in the "far wind" region, $R_{\rm min} > 
\Gamma$. For low $R_{\rm min}$, close to the light cylinder, the pulse shape is almost symmetrical with respect to its peak (also seen in Fig.~\ref{sample}). As one moves towards larger 
$R_{\rm min}$, the pulses stop being symmetrical and instead present an abrupt rise in luminosity followed by a gradual fall, making them asymmetrical with respect to the peak. Also, the 
total luminosity in the pulse is reduced by many orders of magnitude as one moves outwards in the wind. In the example of Fig.~\ref{Rmin} a change from $R_{\rm min}=1$ to $R_{\rm 
min}=30$ results in the decrease in luminosity by seven orders of magnitude. These results agree with previous investigations of pulses from the far wind region (for example \cite{kirketal02}).

From the sky maps we have shown we can conclude that the separation of the two pulses depends on $\chi$ and $\zeta$, with a separation of $\Delta\phi \sim 0.5$ when the line of sight
lies at the equatorial plane. However, as we have pointed out above, the phase-averaged luminosity changes significantly only
if the line of sight lies above the wind, i.e. if $\zeta < \pi/2 - \chi$. In the opposite case the overall luminosity is not sensitive to variations of $\chi$ and $\zeta$.

\subsection{Phase-averaged spectra}

\begin{figure}
\begin{center}
\includegraphics[scale=0.9]{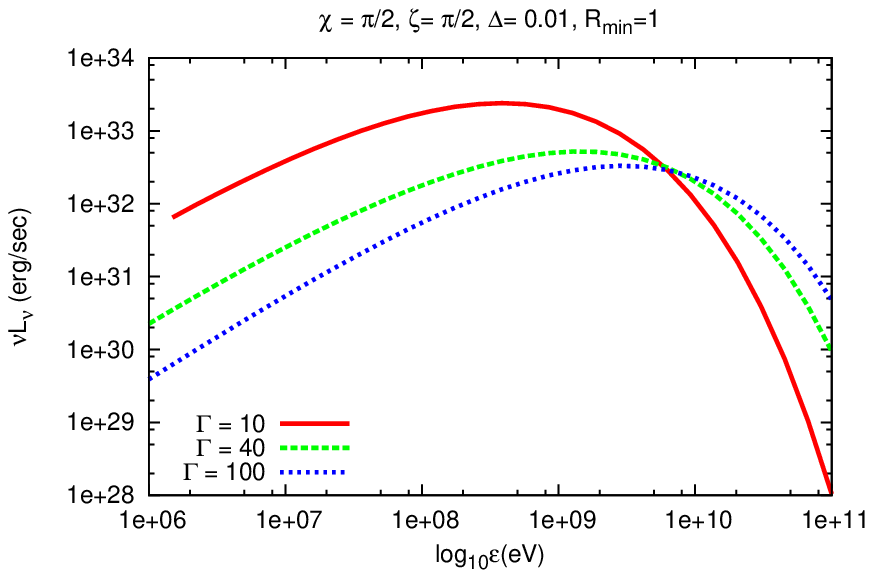}
\includegraphics[scale=0.9]{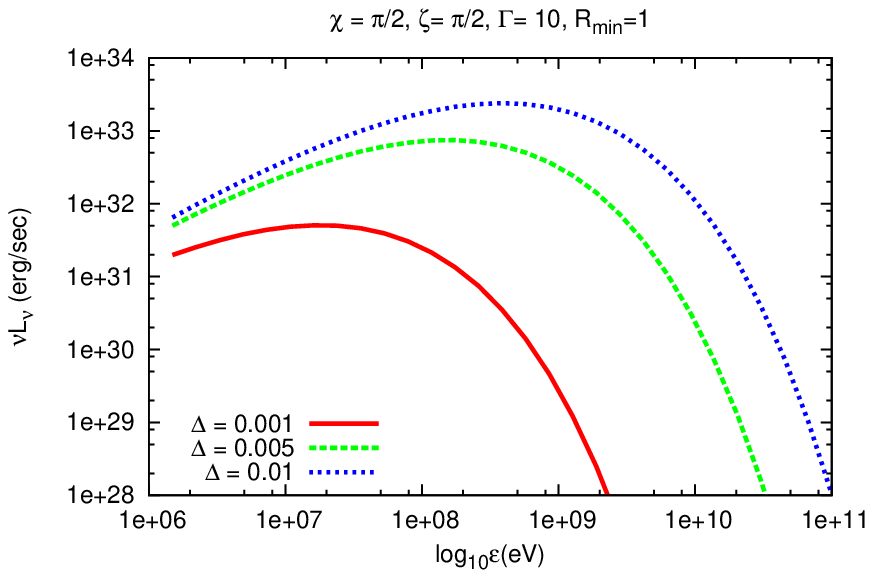}
\includegraphics[scale=0.9]{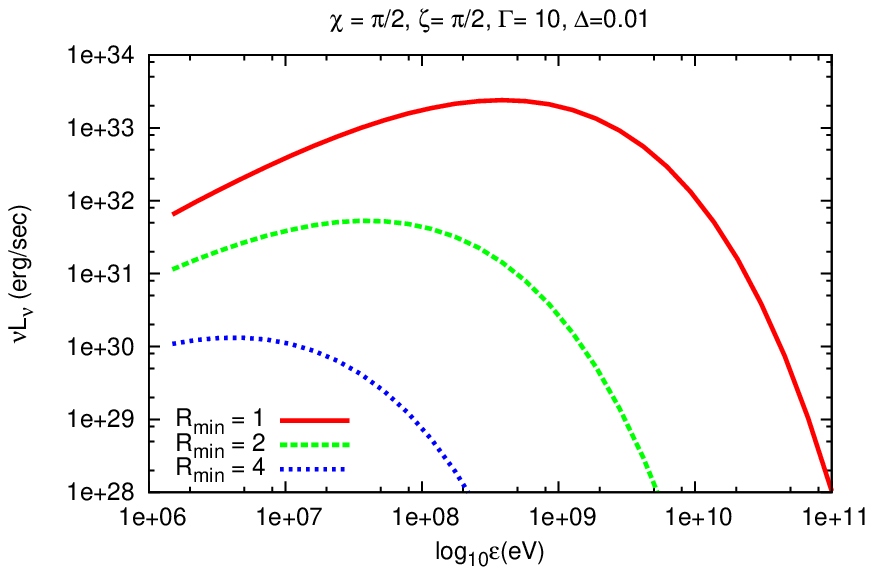}
\end{center}
\caption{Phase-averaged spectra for the same model pulsar that was used in Figs.~\ref{sample} and \ref{Rmin}. All the plots have been made using $\chi = \pi/2$ (perpendicular rotator) and $\zeta 
= \pi/2$. Here the variation of the phase-average flux with the parameters $\Gamma$, $\Delta$ and $R_{\rm min}$ is shown.}
\label{spectra}
\end{figure}

In Fig.~\ref{spectra} we have used the same model pulsar as in the previous section to calculate phase averaged spectra and their dependence on the model parameters. In the upper plot 
we show the variation of the spectrum when changing the bulk Lorentz factor $\Gamma$. As is expected from Eq.~\ref{cutoff}, the peak of the spectrum rises linearly with  $\Gamma$, 
however the overall luminosity falls with rising $\Gamma$. This can be attributed to the diminishing of the area of the sheet from which doppler-boosted radiation is received: the 
boosted area scales as $1/\Gamma^2$, something that is only partly compensated for by the square of the doppler factor in the calculation of the received flux (given in the Appendix). The 
most strongly boosted radiation comes from the line of sight, however the luminosity of the region directly intersecting the line of sight is not the major contribution to the overall observed 
luminosity. This is caused by the fact that one observes only the effects of the perpendicular (i.e. azimuthal) field at $\vartheta = \zeta$, while the effects of the much larger poloidal field 
$B_r$ come from a region at angle $\sim 1/\Gamma$ to the line of sight. Therefore when the line of sight is directly aligned to the edge of the sheet, where $\zeta = \pi/2-\chi$, still two 
pulses of the same amplitude are observed, coming from the two parts of the folded current sheet that are at angle $\sim 1/\Gamma$ to the line of sight, whereas the edge itself contributes 
only a little to the overall flux. This is different to what has been predicted in the past \citep{petri11}. However, a single wide pulse will be observed when looking {\em 
over} the edge of the sheet, the disadvantage in this case being that the luminosity decreases quickly with decreasing $\zeta$ making such pulses difficult to detect.
 
In the middle panel the change in the phase-averaged spectrum is shown when varying $\Delta$. The dramatic rise of the cutoff energy and of the luminosity with $\Delta$ can be explained 
by the dependency of the peak of the spectrum on $\Delta$, given in Eq.~\ref{cutoff}, as well as the rise of the relativistic temperature in the current sheet with $\Delta$ as seen in 
Eq.~\ref{temperature}.

In the lower panel we demonstrate the dependence of the cutoff and luminosity on the minimum radius $\Rmin$. The peak frequency of the spectrum and the overall luminosity 
both rise very strongly with decreasing $\Rmin$, something that is in agreement with Eq.~\ref{cutoff} and is also expected because of the dependence of the temperature of the sheet
particles on radius, given in Eq.~\ref{temperature}. Therefore the bulk of the received high-energy radiation comes from a region of limited range in $R$ close to 
$\Rmin$. 

\subsection{Examples}

\begin{figure*}
\begin{center}
\includegraphics[scale=0.9]{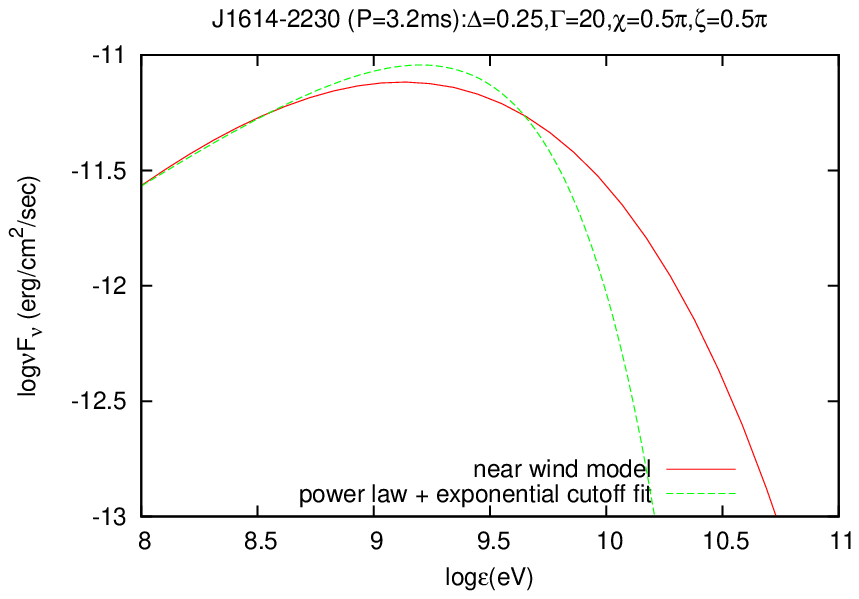}
\includegraphics[scale=0.9]{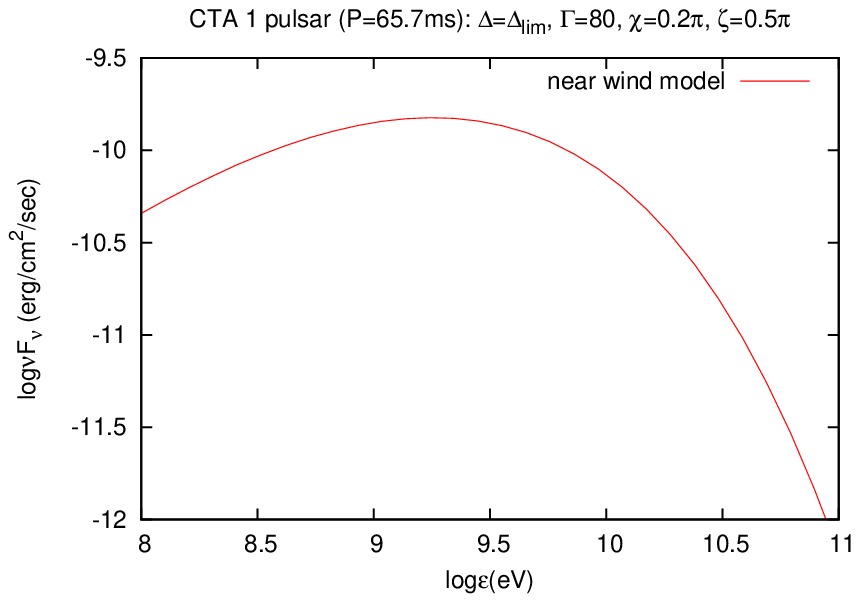}
\includegraphics[scale=0.9]{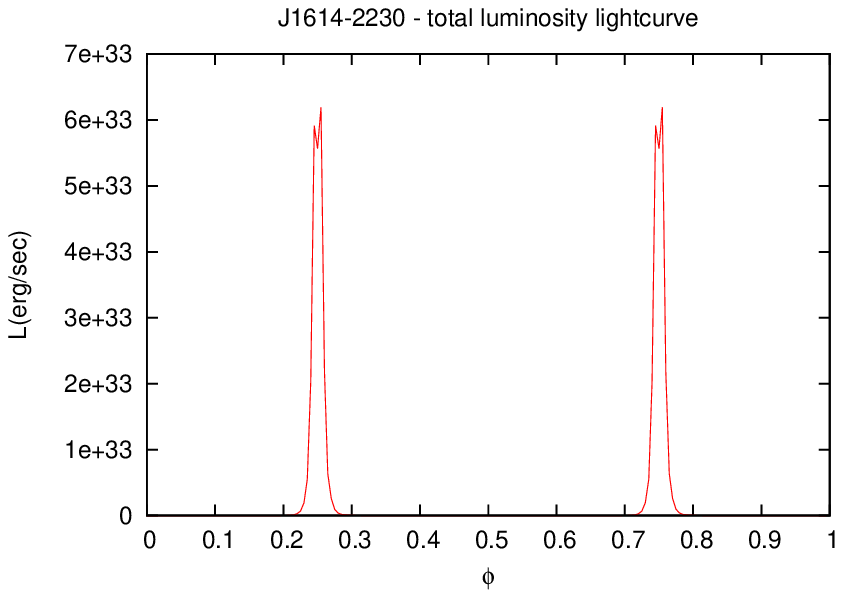}
\includegraphics[scale=0.9]{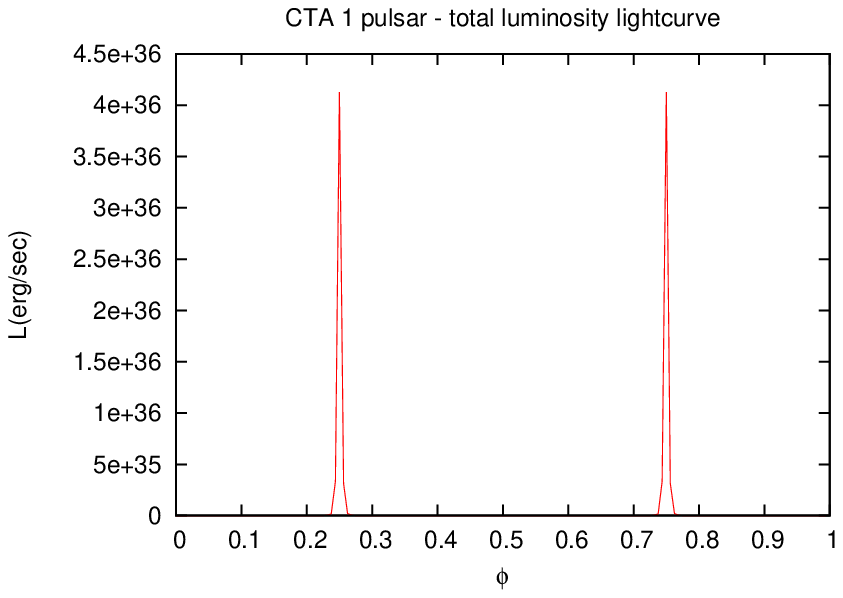}
\end{center}
\caption{The phase-averaged spectrum and simulated lightrcurve of millisecond pulsar J1614-2230 as calculated using our model, along with the best power-law plus exponential cutoff fit 
\citep{abdoetal09}. The parameters that were used are $\Delta = 0.25$, $\Gamma = 20$, $\chi = \zeta = \pi/2$ for the millisecond pulsar (left). For the young pulsar in the CTA 1 
supernova remnant we used $\Delta = \Delta_{\rm lim}$, $\Gamma = 80$, $\chi = \pi/5$ and $\zeta = \pi/2$.}
\label{realpulsars}
\end{figure*}

In Fig.~\ref{realpulsars} we present two examples of spectra and lightcurves of {\em Fermi}/LAT detected pulsars, calculated using our model. The first example is of the millisecond pulsar PSR 
J1614-2230, which has a period of $P=3.2$ms and a spindown 
luminosity of $\Edot = 5$. Its spectrum calculated according to our model can be seen in the upper left panel of Fig.~\ref{realpulsars}, along with the best fit power law with exponential cutoff, 
as given in \cite{abdoetal09}. The parameters used were $\Delta = 0.25 $, $\Gamma = 20$, $\chi =\pi/2$ and $\zeta = \pi/2$. The lightcurve  corresponding to the same parameters is 
seen in the lower left panel.  The phase of the two peaks is correctly predicted, however the width of the simulated pulses is narrower than what is observed. This is because in this example
$\Delta >  1/\Gamma$, and in this case the lightcurve cannot be accurately predicted by our model (as explained in Appendix~\ref{apppend2}).

The second example, shown in the plots on the right of Fig.~\ref{realpulsars}, is of pulsar PSR J0205+6449, which was discovered by {\em Fermi}/LAT in the supernova remnant CTA1. This pulsar 
has a high spindown luminosity of $\Edot = 2.7 \times 10^4$ and a period of $P = 65.7$ms. It is a young, energetic pulsar for which $\Delta_{\rm lim} \simeq 1.5\times 10^{-3}$. The 
parameters that we have used for these plots are $\Delta=\Delta_{\rm lim}$, $\Gamma=80$, $\chi = \pi/5$ and $\zeta = \pi/2$. The relatively high Lorentz factor needed to reach the 
cutoff of $\mathcal{E} \approx 3 \rm GeV$ causes the pulses to be rather sharp, something that is also observed with {\em Fermi}/LAT. 

From these two examples as well as from the phase-averaged spectra shown in Fig.~\ref{spectra} it can be seen that our model predicts a less steep fall and a substantially larger energy flux 
at photon energies above 10 GeV than would be expected using a power-law plus exponential cutoff fit, something that implies that for the more energetic pulsars, which have larger bulk 
Lorentz factors, one could detect radiation in the sub-TeV to TeV regime explored by Cerenkov telescope arrays. Specifically a good fit to our spectra is given by a power-law with a 
sub-exponential cutoff:
\begin{equation}
\frac{dN}{d\mathcal{E}} \propto \mathcal{E}^{-p} \exp\left[ - \left( \frac{\mathcal{E}}{\mathcal{E}_{\rm cutoff}} \right)^b\right]
\end{equation}Our model predicts a cutoff with $b = 0.35$. This index is not sensitive to the obliquity $\chi$ or the observer angle $\zeta$, as long as the line of sight intersects the 
equatorial current sheet, i.e. when $\zeta > \pi/2 - \chi$. The index $b$ can reach the value $\sim 0.4$ for $\zeta < \pi/2 - \chi$, however the very low luminosities that are expected
in that case make the majority of those objects unobservable.

\subsection{Peak energies and $\gamma$-ray luminosities}

\begin{figure*}
\includegraphics[scale=1.]{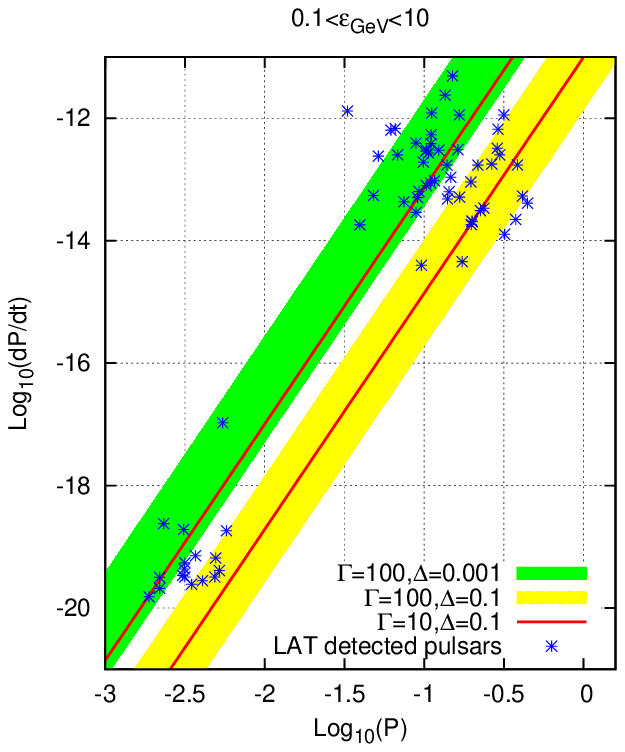}
\includegraphics[scale=1.]{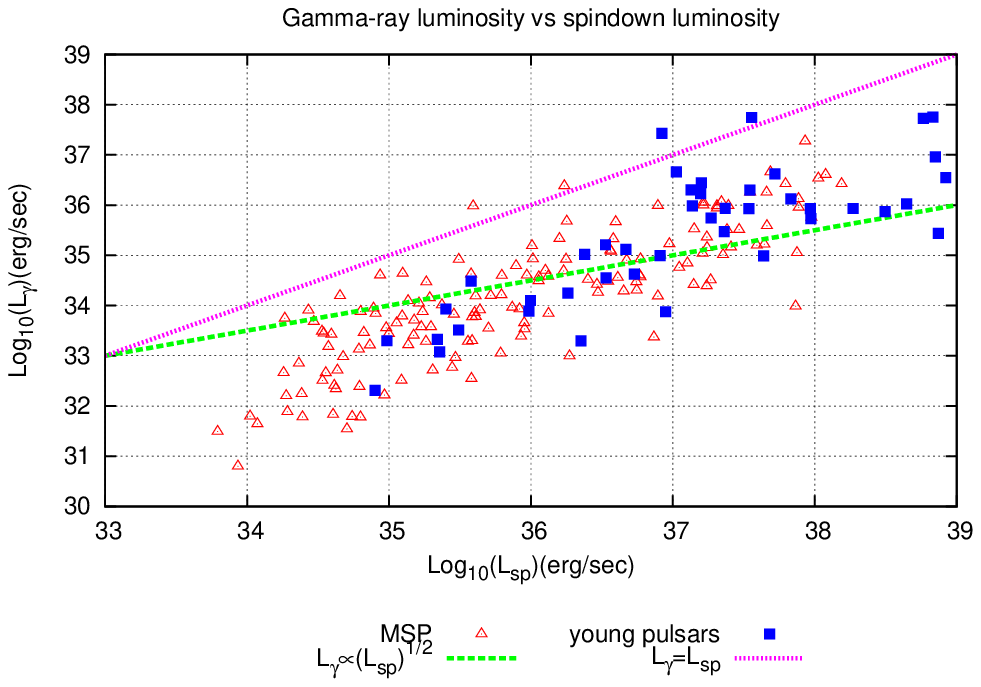}
\caption{On the left: a $P-\dot{P}$ diagram including the pulsars observed by {\em Fermi}/LAT and a plot of the predicted $\gamma$-ray luminosity in the range 0.1-10 GeV for a random sample 
of pulsars. On the right: the gamma-ray luminosity $L_{\gamma}$ versus the spindown luminosity $L_{\rm sp}$ for two random samples of millisecond and young pulsars. The lines $L_{ 
\gamma} = L_{\rm sp}$ and $L_{\gamma} \propto L_{\rm sp}^{1/2}$ are also shown.}
\label{ppdot-lgamma}
\end{figure*}

In Fig.~\ref{ppdot-lgamma} we show a prediction of the region in the $P-\dot{P}$ where spectra that peak in the range $0.1-10 \, \rm GeV$ are espected, according to our 
model. Two 
shaded regions are shown that correspond to $\Delta = 0.1$ and $\Gamma=100$ (yellow) and $\Delta = 0.001$ and $\Gamma = 100$ (green). We have also plotted the lower and upper 
borders 
of the region that corresponds to $\Delta = 0.1$ and $\Gamma = 10$ (red lines). It is obvious that almost all pulsars already detected by {\em Fermi}/LAT fall in at least one of these
regions, leading to the conclusion that one can accommodate the cutoffs of almost all pulsars using parameters in the range $0.001 \leq \Delta \leq 0.1$ and $10 < \Gamma < 100$.

In principle the peak energies emitted by the near wind region of a pulsar wind are not constrained to the GeV regime. Peaks at TeV energies are likely excluded by the synchrotron limit given
in Eq.~\ref{max1}, however there should be a population of pulsars with spectra peaking in the MeV regime (or lower). These pulsars can be found in a region of the $P-\dot{P}$ diagram 
roughly below the yellow shaded region of the $P-\dot{P}$ diagram in Fig.~\ref{ppdot-lgamma}. The population of MeV pulsars will include objects with 
spindown luminosities lower than the ones which are detected by {\em Fermi}/LAT. This becomes obvious by inspecting Eq.~\ref{cutoff}: for lower spindown luminosities the same 
$\Gamma$ and $\Delta$ lead to lower peak energies. For example a pulsar which is a perpendicular rotator with $P=0.5\: s$ and $\dot{P} = 10^{-15}$ with bulk Lorentz factor 
$\Gamma=10$ and  $\Delta = 0.1$ will radiate a spectrum that peaks at energies $\mathcal{E} \simeq 0.3 \: \rm MeV$.

The overall luminosity emitted by the near region of a pulsar wind from a perpendicular rotator according to our model can be estimated to within a factor of two by the expression:
\begin{equation}
\mathcal{L}_{\rm tot} \simeq \frac{2.7\times 10^{31}}{\Gamma} \left( \frac{\blc}{10^3}\right)^{4.7} \left( \frac{P}{0.1} \right)^{3.77} \left( \frac{\Delta}{0.1} \right)^{1.7}
\label{gammaluminosity}
\end{equation}where $\blc$ is given in G and $P$ in s. In terms of spindown luminosity the above can be expressed as:
\begin{equation}
\mathcal{L}_{\rm tot} = \frac{7.6\times 10^{29}}{\Gamma} \left( \frac{P}{0.1} \right)^{-0.9} \Edot^{-2.35} \left( \frac{\Delta}{0.1} \right)^{1.7}
\end{equation}It is obvious from the above expression that, since no trend is observed for $\gamma$-ray pulsars to have a $\gamma$-ray luminosity that is anti-correlated to their spindown 
luminosity, the values of $\Gamma$ and $\Delta$ should vary significantly from object to object.

\begin{figure}
\begin{center}
\includegraphics[scale=0.6]{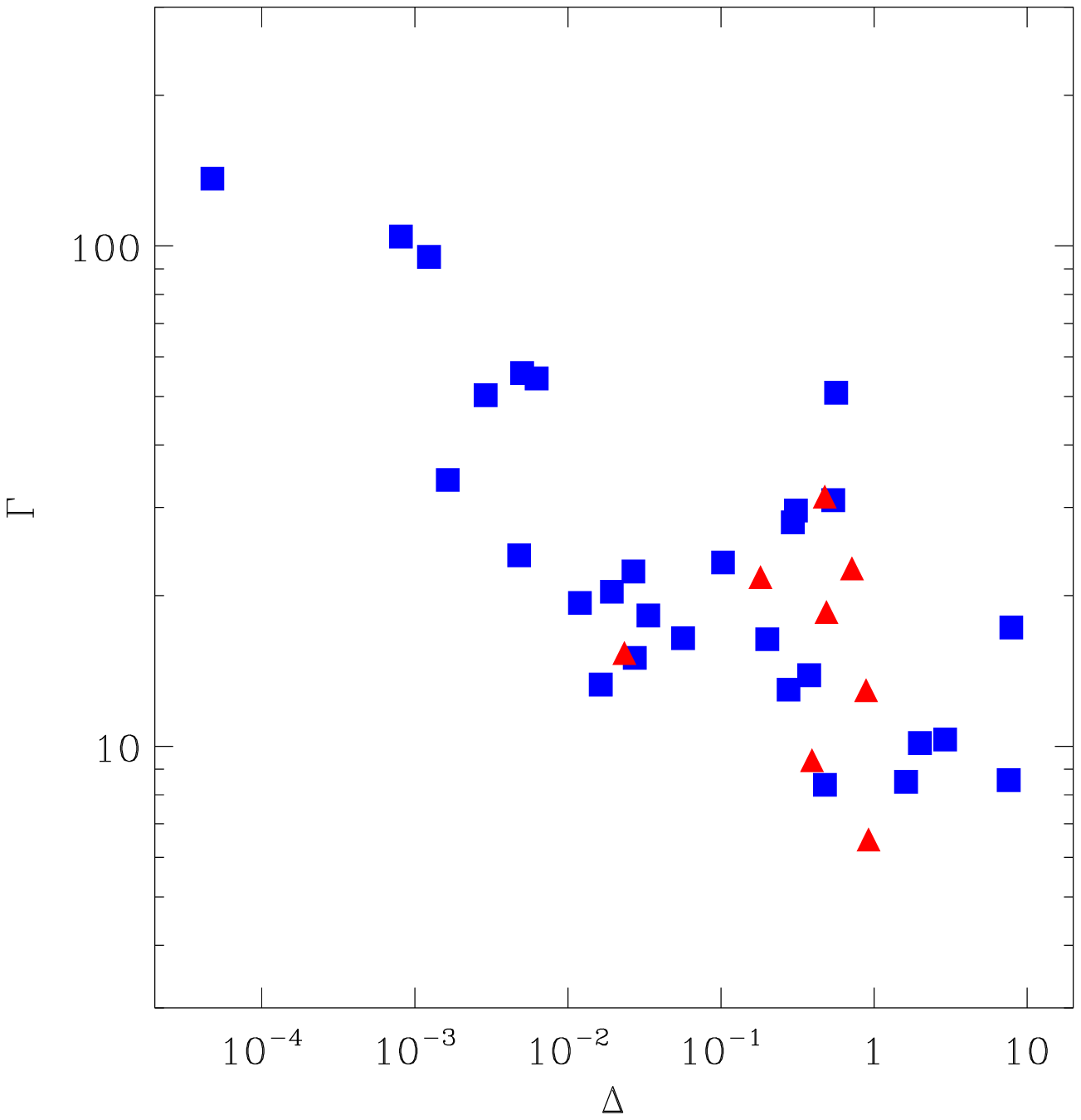}
\end{center}
\caption{The parameters $\Delta$ and $\Gamma$ estimated for the pulsars in the {\em Fermi}/LAT one-year catalog. Blue squares are the young pulsars while red triangles represent the 
millisecond pulsars.}
\label{DG}
\end{figure}

If the peak of the spectrum falls into the $\gamma$-ray regime, then Eq.~\ref{gammaluminosity} gives a good estimate of the $\gamma$-ray 
luminosity of the pulsar. The dependence of $\mathcal{L}_{\rm tot}$ on $\chi$ and $\zeta$ is very weak as long as the line of sight intersects the wind, $\zeta > \pi/2 - \chi$, therefore one
can use the above formula to reach approximate conclusions about the $\gamma$-ray luminosity of a pulsar. Eqs.~\ref{gammaluminosity} and \ref{cutoff} (which gives us an estimate
of the peak energy of the phase-averaged spectrum) show that, once $P$ and $\dot{P}$ are known, the emitted luminosity and peak energy can be fixed essentially by the two parameters 
$\Delta$ and $\Gamma$. These two equations can also be inverted to deduce values of $\Delta$ and $\Gamma$ for a sample of pulsars for which $P$, $\dot{P}$, $\mathcal{L}_{\gamma}$ and 
$\mathcal{E}_{\rm GeV}$
are known. This was done for the pulsars of the First {\em Fermi}/LAT pulsar catalog and the results can be seen in Fig.~\ref{DG}, where the $\Delta$ and 
$\Gamma$ have been plotted for each object of the sample. 

The blue squares represent the young pulsars in the catalog while the red triangles are the millisecond pulsars. For most objects $\Delta$ ranges between $10^{-3}$ and 1, and a trend is
observed for the millisecond pulsars to have larger values of $\Delta$, slightly smaller than unity. The bulk Lorentz factor $\Gamma$ ranges roughly between 10 and 100, with most 
objects having relatively low Lorentz factors in the range 10-30. For the most part the deduced $\Gamma$ and $\Delta$ are reasonable, except for a region where values $\Delta > 1$ are 
predicted, which are unphysical since $\Delta$ has to be less
than unity for the current sheet to be narrower than the wind's half wavelength. This is a result of the fact that the $\gamma$-ray luminosity of these objects is 
relatively high, so that by Eq.~\ref{gammaluminosity} a higher $\Delta$ is needed. Our model is possibly too simple to accomodate all of the pulsars in the catalog, and maybe there is 
a contribution from another physical mechanism or emission region which results in higher luminosities in the exceptional cases of the objects with $\Delta >1$.

In the diagram on the right of Fig.~\ref{ppdot-lgamma} we have plotted the $\gamma$-ray luminosity $L_{\gamma}$ of two random samples of pulsars (millisecond and young pulsars) with
respect to their spindown luminosity 
$L_{\rm sp}$. To this purpose two random samples of pulsars were used, with periods and period derivatives in the ranges $-2 < \log P < 0$ and $-15 < \log\dot{P} < -11$ (for the young 
pulsars) and $-3 < \log P < -2$ and $-20 < \log\dot{P}<-17$ (for the millisecond pulsars). We have chosen $\zeta$ and $\chi$ randomly in the interval $[0,\pi/2]$, the bulk Lorentz factor 
in the interval $1 < \log\Gamma < 2$ and the sheet thickness either within $\log\Delta_{\rm lim} -2 < \log\Delta < \log\Delta_{\rm lim} $ in the case $\Delta_{\rm lim} < 0.1$ or within 
$-3 < \log\Delta < -1$ if $\Delta_{\rm lim} > 0.1$. The $\gamma$-ray luminosity was calculated by integrating the phase-averaged flux between 100 MeV and 100 GeV and using
a beam correction factor $\mathcal{F}_{\Omega}=1$, in order to have a direct comparison with the results of \cite{latfirstyear}.

About  12\% of all young pulsars in our sample and 30\% of the millisecond pulsars have a cutoff within the inteval 100 MeV - 10 GeV and are shown in Fig.~\ref{ppdot-lgamma}. These percentages
reflect our choice of the range in $\Delta$. We have constrained $\Delta$ to be within two order of magnitudes of the value $\min(\Delta_{\rm min},0.1)$. If the range of $\Delta$ were narrower, 
resulting
in systematically larger values, then the percentages quoted above would rise. Lower choices of $\Delta$ lead to lower $\mathcal{E}_{\rm GeV}$ so there are more pulsars that have peaks below
100 MeV and therefore don't make the cut.
 
The large 
beam correction factor used for this plot has as a result the occurrence in some rare cases of objects with $L_{\gamma} > L_{\rm sp}$, whereas in most cases the $\gamma$-ray luminosity is
a few percent of $L_{\rm sp}$. The young pulsars tend to lie higher in the diagram than millisecond pulsars, something that reflects their larger spindown luminosity in combination with 
the fact that the gamma-ray efficiency is similar in both samples. 

In \cite{latfirstyear} a trend was reported that the gamma-ray luminosity is proportional to the square root of the spindown luminosity for the highest luminosities: $L_{\gamma} \propto
L_{\rm sp}^{1/2}$. We don't have a clear indication of such an effect in our simulated sample, meaning that the gamma-ray efficiency of the highest spindown pulsars is predicted to be 
similar to the one of the lower spindown ones. The best fit line to the points in Fig.~\ref{ppdot-lgamma} gives $L_{\gamma} \propto L_{\rm sp}^{0.95}$ for both young and millisecond pulsars
and if the statistical error is taken into account, then the relationship is compatible with the proportionality $L_{\gamma} \propto L_{\rm sp}$.

\section{Discussion and conclusions}

In this paper we have shown how gamma-ray pulses can naturally arise within the framework of the pulsar wind's equatorial current sheet outside (but close to) the light cylinder. 
The advantage of this emission model is that it is an intrinsic mechanism that naturally produces peak energies in the MeV-GeV range. It can give meaningful results 
for almost all pulsars, irrelevant of age or environment, employing only a few parameters: the bulk Lorentz factor of the outflow $\Gamma$, the sheet thickness $\Delta$,  the obliquity $\chi$ 
and the angle of the rotational axis to the line of sight $\zeta$. To these the minimum radius $R_{\rm min}$ can be added (which, in the calculated examples, has been set equal to 
$R_{\rm min}=1$). However it is important to note that $R_{\rm min}$ has to be sufficiently close to the light cylinder, so that the peak energy of the spectrum reaches the GeV regime.
The parameters of the model are constrained by the characteristics of the pulsar, and by the physics of the current sheet. The novelty of our model is that it makes an attempt to 
account for the previously ignored region of the wind between $R=1$ and $R\simeq \Gwind$, and is therefore able to take advantage of the strong poloidal field close to the light cylinder, 
something that has not been discussed in previous wind models. 

The predictions of our model are the following:
\begin{enumerate}
\item Two pulses per pulsar period are expected. The pulses have the same amplitude.
\item The shape of the pulses is symmetric with respect to their peak, as long as emission starts close to the light cylinder, and become increasingly asymmetric with rising $R_{\rm min}$.
\item There is no significant interpulse emission in the GeV range.
\item The width of the pulse decreases with increasing photon energy.
\item The separation of the peaks varies with the obliquity $\chi$ and the angle to the line of sight $\zeta$, as in previous models of pulsed radiation from the pulsar wind.
\end{enumerate}

The features of the pulsed emission that cannot be explained by our model, such as the interpulse or the peaks of different intensity, could be accomodated by a model in which 
radiation comes both from within the light cylinder, possibly in an outer gap, as well as from the beginning of the wind. The difficulty of outer gap models to produce double peaked lightcurves, 
as
noted in \cite{bai+spitkovsky10}, in combination with the difficulty of our near wind model to produce single pulses, point to the need for a combined model in which the high energy emission 
starts within the light cylinder and continues in the equatorial current sheet outside the light cylinder. Therefore the physics of the region of formation of the current sheet is very important in the
endeavour to understand $\gamma$-ray emission from pulsars.

For pulsars in binaries, the possibility exists that the favoured radiation process is not synchrotron but rather inverse Compton on the low frequency photons of the 
massive star, as investigated in \cite{petri+dubus11}. In this case the peak energies and emitted luminosities by the two mechanisms can be comparable, depending on whether the comoving
energy density of the low frequency photons in the wind frame is comparable to the magnetic field energy density close to the light cylinder. For the pulsar B1259-63, which is a member of a 
binary with a B2e star, synchrotron radiation close to the light cylinder would give a peak energy $\mathcal{E}_{\rm GeV}\sim 3$ (assuming $\Gamma=10$ and $\Delta =0.01$), while inverse 
Compton radiation on the star's photon gives a peak energy that is constrained by the Klein-Nishina limit to $\mathcal{E}_{\rm GeV}\sim 10$. In this case the low frequency photon energy density is 
several orders of magnitude lower than the one of the magnetic field, therefore synchrotron radiation should be the dominant component. This, however, is not necessarily the case
for other binary systems, such as LS 5039, where the seed photon energy density exceeds the one in PSR B1259-63 by a factor of 200. The detection of orbital modulation of 
the received $\gamma$-ray flux can be the criterion on which to distinguish between the two emission mechanisms.

A possible issue with our model is that the supersonic solution of \cite{bogovalov99} might not apply for the objects of lower bulk Lorentz factors. Particularly, the wind can accelerate, as 
is predicted in various magnetohydrodynamic models of pulsar winds \citep{beskinetal98,kirketal09}. In this case the shape of the current sheet can in reality be different, since there is generally also 
a polar 
magnetic field component, which in our model was zero. However, as was mentioned above, the very steep dependence of $\mathcal{E}_{\rm GeV}$ on $R$ implies that 
even if the parameters $\Delta$ and $\Gamma$ changed with $R$, the present model would still be able to explain the main features of the observed spectra. Adding to these considerations the fact 
that the particle density within the current sheet and the magnetic field decrease as the wind expands to larger $R$, the conclusion is reached that the emitted luminosity, which depends on 
$N'_0$ and $B'_0$ will also decrease with 
radius, and this in combination with the decreasing peak of the spectrum means that the main contribution to the 
spectrum near the cutoff will still come from a very limited region close to $\Rmin$. Therefore, possible changes in the dynamic evolution of the wind should not have a large effect to the observed
spectrum and luminosity, while the exact shape of the current sheet should influence mainly the shape of the pulse, but not the cutoff energy or the emitted luminosity.

It is expected that reconnection will generally alter the physics of the current sheet beyond the pulsar's light cylinder, and might result in an evolution of the 
current sheet thickness and the wind Lorentz factor, especially in objects for which the radiative timescales of the thermal particles are very short in comparison
to the evolution timescale of the magnetic field. Furthermore, particle acceleration takes place during the reconnection process which might result in a non-thermal tail to the thermal distribution, 
the characteristics of which will change as the current sheet evolves with radius. These phenomena will leave their imprint on 
the pulsed high-energy spectrum, and need to be investigated in a self-consistent way, something that is beyond the scope of this article. We can, however, give a
simple order-of-magnitude argument about the reconnection-accelerated particles: deep in the current sheet particles can be accelerated by reconnection electric 
fields to energies higher than the thermal peak. These particles would have 
Lorentz factors extending to $\gamma \sim \xi \alc/\Gamma$ (in the wind frame), thus giving rise to photons up to an energy
\begin{equation}
\mathcal{E}_{\rm GeV}\sim 8.8 \frac{\xi^2 \Edot^{3/2}}{\Gamma P}
\end{equation}where $\xi = E'/B'_0 <1$. This could extend beyond the thermal peak for large enough $\xi$ or $\Edot$, whereas millisecond pulsars are again favoured by the 
dependence on the inverse of the period.
If these high energy non-thermal particles escape the acceleration site and radiate in the field within the current sheet, their emission could give rise to a power-law 
tail extending beyond the GeV cutoff \citep{zenitani+hoshino08}. This 
mechanism should be prominent for objects for which $t'_{\rm s}/t'_{\rm R}<1$, which tend to cluster at the upper left part of the $P-\dot{P}$ diagram, as discussed. It is for these pulsars that 
acceleration by reconnection could become prominent and be observed in the form of power-law tails in the GeV-TeV regime. 

Another possibility is that the particles in the current sheet are already accelerated
to a power-law distribution when the sheet starts radiating, in which case a different distribution function would have to be used in order to calculate the current sheet parameters
\citep{balikhin08}. This might apply particularly to millisecond pulsars, which have lower surface magnetic fields which lead to lower pair production rates, and therefore less dense plasmas in 
their magnetospheres. In this case a non-thermal particle distribution seems likely to describe the physics of the current sheet in a more consistent way. We defer the investigation of such 
particle distributions to a future article. 
\\

\begin{acknowledgements} I.A. would like to acknowledge support from the EC via contract ERC-StG-200911. I.A. would like to thank J\'er\^ome P\'etri, Gilles Henri and Anatoly Spitkovsky for useful 
discussions. 
\end{acknowledgements}

\begin{appendix}

\section{Calculation of received synchrotron spectrum by a relativistcally moving sheet}
\label{append1}

The flux that an observer receives from a relativistically moving source is calculated by the formula \citep{lind+blandford85}:
\begin{equation}
F_{\nu} = \frac{1}{d^2} \int \mathcal{D}^2 j'\left( \nu' \right) dV
\end{equation}
\begin{itemize}
\item Primed quantities are in the wind frame, i.e. the frame where the outflow is at rest. Since the wind is assumed to be flowing radially, at each azimuth and polar angle ($\varphi,
\vartheta$)
there is a different, local, radially moving "wind frame". Quantities pertaining to a point of the wind with coordinates ($r,\vartheta,\varphi$) are calculated in this {\em local} wind frame. 
\item When calculating the 
emission coefficient $j'$ we should take care to calculate it in the direction {\em of the line of sight, taking into account the aberration of photons} in the wind frame (we will return to this matter 
later).
\item The volume element is taken in the observer's frame (or the frame in which the pulsar is at rest) and is $dV = r^2 \sin\vartheta dr d\vartheta d\varphi$.
\item $\mathcal{D}$ is the doppler factor, which depends on the angle between the line of sight and the direction of motion of the wind. It is calculated by the expression
\begin{equation}
\mathcal{D}  = \frac{1}{\Gamma (1-\beta \hat{O}\cdot\hat{r})}
\label{dopplerfactor}
\end{equation}
The unit vector $\hat{r}$ is the radial unit vector in a spherical coordinate system, the center of which is located at the pulsar. $\hat{O}$ is a unit vector in the direction of 
the observer. The cosine of the angle between
these two vectors depends on $\vartheta$ and $\varphi$, and therefore $\mathcal{D}=\mathcal{D}(\vartheta,\varphi)$. The observed frequency is $\nu = \mathcal{D} \nu'$.
\item We will assume that the direction to the line of sight is at polar angle $\zeta$ and at $\varphi = 0$. In a cartesian system of coordinates, then, we have: 
$$\hat{O} =  \sin\zeta \hat{x} + \cos \zeta \hat{z} $$whereas the radial unit vector can be expressed as: $$ \hat{r} = \sin\vartheta \cos\varphi \hat{x}+\sin\vartheta\sin\varphi 
\hat{y}+\cos\vartheta
\hat{z} $$which gives us: \begin{equation}
\hat{O}\cdot\hat{r} = \sin\zeta\sin\vartheta\cos\varphi + \cos\zeta \cos\vartheta
\label{odotr}
\end{equation}
\item Finally, $d$ is the distance of the object from the observer. We define a luminosity associated with the pulsar as:
\begin{equation}
\mathcal{L}_{\nu} = 4\pi d^2 \mathcal{F}_{\Omega} F_{\nu} = 4\pi \mathcal{F}_{\Omega} \int \mathcal{D}^2 j'\left( \nu' \right) d^3x
\label{luminosity}
\end{equation}The parameter $\mathcal{F}_{\Omega}$ is the beam correction factor, which is taken to be equal to  $\mathcal{F}_{\Omega}=(4\pi)^{-1}$ in the plots of the pulsar lightcurves and 
spectra, Figs.~\ref{sample}, \ref{Rmin} and \ref{spectra}. It is, however, taken to be equal to unity $\mathcal{F}_{\Omega}=1$ for the $L_{\gamma}-L_{\rm sp}$ plot in Fig.~\ref{ppdot-lgamma}.
\end{itemize}
\end{appendix}

\begin{appendix}

\section{Calculation of the emission coefficient}
\label{apppend2}

The emission coefficient can be calculated by multiplying the energy distribution of the particles by the single-electron synchrotron spectrum. This implies that the electrons are 
relativistic, which is, strictly speaking, not true for the whole population, since the distribution starts from $\gamma'=1$. However, since the temperature of the distribution is relativistic this 
approximation will 
not introduce a significant error. The distribution gives the number of electrons per unit gamma factor, per solid angle and per unit volume:
$$ \frac{dN'}{d^3x'd\gamma'd\Omega'}= \frac{N'_0}{4 \pi \Theta K_2(1/\Theta)} 
\gamma' \sqrt{\gamma'^2-1} e^{-\gamma'/\Theta} \cosh^{-2} \left(\frac{X'}{\Delta_{X'} }
\right)$$where $\Theta \gg 1$ is the dimensionless 
temperature of the distribution in units of $mc^2$ as defined in the text. This distribution is isotropic and its density falls towards the edge of the sheet.
The direction perpendicular to the sheet midplane is denoted by $X'$ and $\Delta_{X'}$ is the width of the sheet at its local rest frame. 

\begin{figure}
\begin{center}
\includegraphics[scale=0.8]{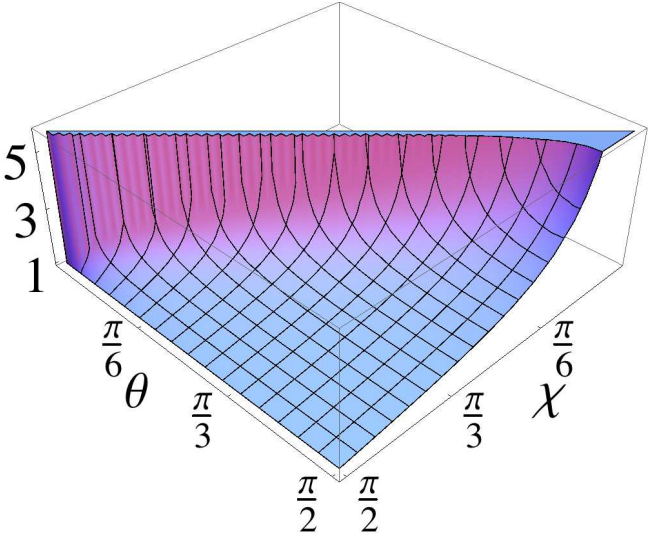}
\end{center}
\caption{The function $C(\vartheta,\chi)$ plotted for $0\leqslant \chi\leqslant \pi/2$,$0\leqslant \vartheta \leqslant \pi/2$. Away from the sheet's edge, where $\vartheta = \pi/2 -\chi$,
its value is close to unity.}
\label{cfunction}
\end{figure}

Using the small argument approximation of the modified Bessel function $K_2$, Eqn.~\ref{k2approx}, the distribution function can be expressed as:
 \begin{equation}
 \frac{dN'}{dV'd\gamma'd\Omega'}= \frac{N'_0}{2 \Theta^3} \gamma' \sqrt{\gamma'^2-1} e^{-\gamma'/\Theta}\cosh^2\left(\frac{X'}{\Delta_{X'}} \right)
 \end{equation}
This then has to be doubled to take account of both species in the sheet (electrons and positrons). The 
single-electron synchrotron spectrum is:
$$ \frac{dE'}{dt' d\nu'} = \frac{\sqrt{3}e^3 B'_{\perp}}{mc^2} F\left( \frac{\nu/\delta}{\nu'_{\rm cr}}\right) \tanh \left( \frac{X'}{\Delta_{X'}}\right) $$ 
$F(x)$ is the well-known synchrotron
function. In the above expression, we encounter the following:
\begin{itemize}
\item The field $B'_{\perp} = B'\sin\alpha$ perpendicular to the direction that the photons which reach the observer travel in the wind frame. $\alpha$ is the angle between
the field in the wind frame and the photons trajectory, which can be computed through the scalar product $\overrightarrow{B'}\cdot \hat{O'}$.
 In order to calculate this, we need to find the 
relativistic aberration of photons which follow the line of sight in the lab frame. This is equivalent to calculating the aberration of the vector $\hat{O}$. The field has only a radial and an 
azimuthal component, so only the aberration of these two components of the unit vector $\hat{O}$ are needed. These components are:
at the point ($r,\vartheta,\varphi$) we have:
\begin{eqnarray}
O_{r} & = & \sin\zeta\sin\vartheta\cos\varphi + \cos\zeta\cos\vartheta \\
O_{\varphi} & = & -\sin\zeta\sin\varphi
\end{eqnarray}The corresponding primed components in the local wind frame are:
\begin{eqnarray}
O'_{r} &=& \frac{O_r - \beta}{1-\beta O_r} \\
O'_{\varphi} &=& \frac{O'_{\varphi}}{\Gamma(1-\beta O_r)}	
\end{eqnarray}and the angle $\alpha$ is \begin{equation} \alpha = \arccos \left( \hat{O'} \cdot \overrightarrow{B'}/B'\right) =  \arccos \left( \frac{O'_r B'_r + O'_{\phi} B'_{\phi}}{\sqrt{(B'_r)^2 + 
(B'_{\phi})^2}}\right)
\label{angle}
\end{equation}while the local magnetic field in the wind frame is: \begin{eqnarray}
B'_r &=& \frac{B_{\rm LC}}{R^2} \\
B'_{\varphi} &=& \frac{B_{\rm LC}}{\beta \Gamma R} \sin\vartheta \\
\end{eqnarray}where the superscript $\rm LC$ refers to quantities at the light cylinder.
\item The ratio $X'/\Delta_{X'}$ can be approximately converted to quantities in the lab frame to give:
\begin{equation}
\frac{X'}{\Delta_{X'}} \simeq \frac{R-R_0(\chi,\vartheta,\varphi,t)}{\Delta}
\end{equation}where $R_0(\chi,\vartheta,\varphi,t)$ is the location of the current sheet point under consideration at time $t$.
\item The critical frequency $\nu'_{\rm cr}$ is
\begin{equation}
\nu'_{\rm cr} = \frac{3 e B'_{\perp}}{4 \pi m c} \gamma'^2 \tanh\left( \frac{X'}{\Delta_{X'}}\right) \label{frequency}
\end{equation}
\item In the general case of the oblique rotator the temperature is given by the expression
\begin{eqnarray}
\Theta  &= &
\Theta_{\perp} \left( \frac{1+\left( \frac{\beta \Gamma}{R_0 \sin\vartheta} \right)^2 }{ 1+\left( \frac{\beta \Gamma}{R_0 \sin\vartheta} \right)^2C(\vartheta,\chi) } \right)^{1/3} \\
C(\vartheta,\chi)& = &1+\left( \frac{\csc\vartheta \cot\chi}{\sqrt{1-\cot^2\chi \cot^2\vartheta}} \right)^2
\end{eqnarray}where $\Theta_{\perp}$ is the value of the temperature for the perpendicular rotator, given in Eqn.~\ref{temperature}.
\end{itemize}
The function $C(\vartheta,\chi)$ is very close to unity for a large range of $\vartheta$ given a $\chi$, and deviates significantly from that value only for values $\vartheta \simeq \pi/2-\chi$, or for small obliquities, as is seen in Fig.~\ref{cfunction}. Therefore, the perpendicular rotator is a good approximation to the general problem, and this is why we have used $\Theta_{\perp}$ in all our estimates.

The emission coefficient $j'$ has to be calculated at the retarded time $$t_{\rm ret} \simeq t - \frac{d}{c} + \frac{\overrightarrow{r}\cdot \hat{O}}{c}$$where $\overrightarrow{r}$
is the position vector of the radiating point in the current sheet, $d$ is the distance of the pulsar from the observer and $t$ is the time that the observer measures. Normalizing 
time to $1/\omega$ and distance to $\rlc$ the retarded time becomes 
\begin{equation} 
t^*_{\rm ret} \simeq t^* - \frac{d}{r_{\rm LC}} + R_0 \hat{r}\cdot\hat{O}
\label{tret}
\end{equation}(normalized time is denoted with an asterisk).
The retarded time enters the computation via the equation for the motion of a current sheet: 
$$ R_0 = \beta \left( \pm \arccos\left( -\cot\chi \cot\vartheta \right) + t^*_{\rm ret} -\varphi \right) $$We replace the retarded time in the last 
equation from \label{tret}, and move the initial condition for time at 
$t^* =d/\rlc$ so that the new observer's time is $t^*_d = t^* - d/\rlc$. We get then an expression for the radius of the current sheet as a function of observer's time,
polar angle and azimuthal angle:
\begin{equation}
R_0 = \frac{\beta \left( \pm  \arccos\left( -\cot\chi \cot\vartheta \right) + t^*_d - \varphi +2 k \pi \right) }{1-\beta \hat{O}\cdot\hat{r}}
\label{R0}
\end{equation}
Depending on how large or small the dot product $\hat{O}\cdot \hat{r}$ is, the radius $R_0$ where the current sheet part under consideration is located, the radiation from which is received at time 
$t^*_d$, can vary strongly. This means that at each moment $t^*_d$ the observer receives radiation that comes from different parts of the current sheet that propagates in the wind, with different 
coordinates $(R_0,\vartheta,\varphi)$. We note here that $R_0$ is considered to be the radius at the midplane of the current sheet.

Finally, the flux can be computed, according to the above, as follows:
\begin{equation}\mathcal{F}_{\nu} = {\rm A}\int \mathcal{D}^2\frac{N'_0}{\Theta^3} \frac{B' \sin\alpha}{B_{\rm cr}} \\ 
F\left(\frac{\nu}{\delta \nu'_{\rm cr}} \right) \frac{\tanh\left(\frac{R-R_0}{\Delta}\right)}{
\cosh^{-2} \left(\frac{R-R_0}{\Delta}\right) }dv d\gamma' \label{flux} \end{equation}
 with the volume element, the function $n(\gamma')$, the constant $A$ and the magnetic field $B'_{0}$ given by:
\begin{eqnarray}
dv & = &R^2 \sin\vartheta dR d\vartheta d\varphi \\
n(\gamma') &=&\gamma'\sqrt{\gamma'^2-1}e^{-\gamma'/\Theta} \\
{\rm A} &=& \frac{\sqrt{3}}{4\pi d^2} \alpha_f mc^2 \rlc^3  \\
B'_0 &=& \frac{B_{\rm LC}}{R_0^2} \sqrt{\left(\frac{R_0\sin\vartheta}{\Gamma}\right)^2+1} \\
\end{eqnarray}
The doppler factor has been given in Eq.~\ref{dopplerfactor} and is calculated with the help of Eq.~\ref{odotr}. The strength parameter $\alc$ and the density in the middle of the current
sheet $N'_0$ are given in the text. The expression for the flux also includes the sine of angle $\alpha$ 
which is calculated by Eq.~\ref{angle}. 
The critical frequency $\nu'_{\rm cr}$ is given in Eq.~\ref{frequency}, with $B'_{\perp} = B'_0 \sin\alpha$. Finally, the above parameters 
are calculated at radius $R_0$ with the help of $t^*_d$ from Eq.~\ref{R0}. The purpose of the integration on 
$R$ is to calculate the integral of the emission coefficient across the sheet, when the sheet midplane is at the point $(R_0,\vartheta,\varphi)$. The numerical computation of the above integral 
for each point in time $t^*_d$, or equivalently each phase in a pulsar period, and each frequency $\nu$, produces the spectra and lightcurves
needed.

Finally, it should be noted that the above prescription for the calculation of the pulsar lightcurves is accurate as long as $\Delta < 1/\Gamma$. In the opposite case, the pulse width is not governed 
by $\Gamma$, but is comparable to $\Delta$, something that is not taken into account in the calculation presented here. However, the phase-averaged fluxes calculated by our model are reliable 
even for $\Delta > 1/\Gamma$.
\end{appendix}

\bibliographystyle{aa}
\bibliography{arka}

\end{document}